# Understanding the Difference between Office Presence and Co-presence in Team Member Interactions

Nils Brede Moe
SINTEF
nils.b.moe@sintef.no

Simen Ulsaker
Telenor
simen.ulsaker@telenor.com

Darja Smite
SINTEF, BTH
darja.smite@bth.se

Jarle Moss Hildrum
Deloite
jhildrum@deloitte.no

Fehime Ceren Ay
Telenor
fehime-ceren.ay@telenor.com

## Abstract

Although the public health emergency related to the coronavirus disease 2019 (COVID-19) pandemic has officially ended, many software developers still work partly from home. Agile teams that coordinate their office time foster a sense of unity, collaboration, and cohesion among team members. In contrast, teams with limited co-presence may experience challenges in establishing psychological safety and developing a cohesive and inclusive team culture, potentially hindering effective communication, knowledge sharing, and trust building. Therefore, the effect of agile team members not being co-located daily must be investigated. We explore the co-presence patterns of 17 agile teams in a large agile telecommunications company whose employees work partly from home. Based on office access card data, we found significant variation in co-presence practices. Some teams exhibited a coordinated approach, ensuring team members are simultaneously present at the office. However, other teams demonstrated fragmented co-presence, with only small subgroups of members meeting in person and the remainder rarely interacting with their team members face-to-face. Thus, high average office presence in the team does not necessarily imply that team members meet often in person at the office. In contrast, non-coordinated teams may have both high average office presence and low frequency of in-person interactions among the members. Our results suggest that the promotion of mere office presence without coordinated co-presence is based on a false assumption that good average attendance levels guarantee frequent personal interactions. These findings carry important implications for research on long-term team dynamics and practice.

## 1. Introduction

Software development companies are, and will increasingly be, places of hybrid working, even those that promote agile values and practices. Agile teams and team members have the flexibility to choose, at least to some degree, between remote and office-based work; in a large-scale agile context, teams probably have different rhythms (Conboy et al., 2023). The documented benefits of flexible working (reduced commute time, better conditions for focused work, and better work–life balance) (Smite, Moe, et al., 2022) have led to many workplaces allowing hybrid working—that is, alternating between working at the office and working remotely (Smite, Christensen, et al., 2022). For teams, hybrid working causes changes in their co-presence rhythms, with members being not quite distributed and not quite co-located but, in the worst case, working from anywhere and intermittently touching base with the office. Whereas hybrid work arrangements in large-scale agile projects have become increasingly common, how to support agile teams that have relied on co-presence and onsite work practices is not yet well understood (de Souza Santos & Ralph, 2022).

Teams in agile software development must be empowered, self-managed, and autonomous: the team decides how to organize the work, including when and how to meet. Orchestrating a project compound of multiple autonomous teams is already a complex task, but when team members exercise individual flexibility and largely work remotely, this task becomes even more complex. This is because a lack of co-presence in the teams significantly affects within-team and cross-team communication as well as the effectiveness of the regular coordination mechanisms when used in a computer-mediated fashion (Espinosa & Carmel, 2003). A recent study also shows that hybrid teams not exercising co-presence experience decreased psychological safety (Tkalich et al., 2022). For successful hybrid working in large-scale agile projects, organizations must find ways to balance individual, team, project, and organizational needs. However, in hybrid working environments, the interests of these groups are in conflict—individual flexibility, productivity, and well-being for team members; effective collaboration, coordination, and spontaneous interaction for teams; and profitability, quality of products and services, employee retention, and attractiveness in the job market for the organization.

Motivated by the need to understand how agile teams handle hybrid working environments, we pose the following research question:

*RQ: How do agile team members organize their work rhythms in a hybrid work environment*?

To answer this question, we conducted a single case study on the patterns of office co-presence among 17 agile teams in a national unit of a large multinational telecommunications operator.

The remainder of this paper is organized as follows. Section 2 presents the background on teamwork and networking in large-scale agile and hybrid software development. Section 3 describes the case companies and the research methodology employed, and Section 4 presents the results. Section 5 discusses the findings, suggests their implications for practice, and presents the study's limitations. Finally, Section 6 concludes the study and proposes avenues for future work.

## 2. Background

### 2.1 Large-scale agile teams

Connections and community are key factors in agile software development. However, these are negatively affected when working remotely. A Microsoft study of over 60,000 employees shows that firm-wide transition to remote work made the collaboration network more static and siloed, with fewer ties that cut across formal business units owing to asynchronous communication (Yang et al., 2022). De Souza Santos and Ralph (2022) studied coordination in hybrid software teams and found a decreasing trend in the feeling of attachment and cohesion in these teams. Several studies have also found that members' interest in collaborative work decreases when the work is done remotely (Technology & reserved, 2021; Smite et al., 2021), and that when joining remote Hackathons they work more in isolation because tasks are split between team members resulting in less collaboration (Moe et al., 2022). In fact, the very future of agile teamwork and work practices was regarded as threatened by hybrid working in light of the increasing coordination challenges and separation of team members (de Souza Santos & Ralph, 2022).

Autonomy is a central principle in agile methods (Ravn et al., 2022). In this work, we rely on the widely used definition of autonomous teams by Guzzo and Dickson (1996) in their review on team performance: "employees that typically perform highly related or interdependent jobs, who are identified and identifiable as a social unit in an organization, and who are given significant authority and responsibility for many aspects of their work, such as planning, scheduling and assigning tasks to members, and making decisions with economic consequence." Based on this definition, the team is supposed to decide the work process.

### 2.2 Networking in large-scale agile software development

Agile software development involves collaboration and communication. In large-scale agile software development, close cooperation and collaboration between all team members and across teams in the organization are essential (Berntzen et al., 2022, 2023); thus, networking is vital in this context. The need for such networking highlights the importance of social capital: "the actual and potential resources embedded within, available through, and derived from the network of relationships possessed by an individual or social unit" (Nahapiet & Ghoshal, 1998). Social capital is key for agile software development: it is a prerequisite for solving problems, making decisions, shifting workload, understanding common goals, and sharing knowledge among or across individuals and teams. Further, access to an informal network is essential when new employees are onboarded (Bauer, 2010; Jones, 1986; Moe et al., 2020).

Networking is essential for novice and mature teams working on complex, unfamiliar, or interdependent tasks owing to the increased need to effectively pull knowledge from outside the team's network (Šmite et al., 2017). The ability to navigate through the organizational network is also an enabler for autonomy (Šmite, Moe, Floryan, et al., 2023) and for sharing knowledge in large-scale agile software development (Smite et al., 2019). Smite et al. (2017) found that in a large-scale distributed setting, network size and networking behavior depend on factors such as company experience, employee turnover, team culture, need for networking, and organizational support. These findings may be relevant in the context of remote and hybrid working as well.

Networking, despite being a prerequisite for large-scale agile software development, is more challenging in distributed teams (Stray & Moe, 2020). When some team members are working from home and others are working from the office, conducting spontaneous discussions is difficult (Tkalich et al., 2022). An extensive study of individual communication networks at Microsoft has revealed the effects of remote work on collaboration networks (Yang et al., 2022); the study suggests that the shift to work-from-home (WFH) has made collaboration networks more heavily siloed across organizational units and more static with fewer new ties. Other researchers have expressed similar concerns regarding the long-term effects of deteriorating social ties (Clear, 2021). However, Šmite, Moe, Klotins, et al., (2023) found that WFH has not significantly affected productivity owing to the strengths of the already established relationships. Accordingly, we argue that networking within an established network is efficient when teams are co-located or distributed, but challenges arise when teams are not synchronized and when new networks need to be created owing to many members working from home, especially the members newly onboarded in a remote or hybrid setting.

## 3. Research methodology

In this study, we exemplify hybrid work rhythms through the individual attendance data and patterns of co-presence of 17 teams partially working from home. We collected data from 24 teams but excluded 7 teams, as they had 4 or fewer members. Ours is a single case study (Yin, 2017) conducted in a national unit of a multinational telecommunications company headquartered in the Nordics. The unit has two offices, where agile software development methods have been employed since 2015. Later, the company underwent another agile transformation to establish a DevOps organization. DevOps is a concept for software development that extends agile principles to the entire software delivery process (Jabbari et al., 2018; Stray et al., 2019) and prescribes structural and procedural changes. This concept emerged from an increasing disconnect between the development and operations functions arising within large software companies (Fitzgerald & Stol, 2017). It is considered a prerequisite for continuous software development, as it enables knowledge sharing and breaks down barriers between development and operations; relies on the automation of build, deployment, and testing systems; and focuses on improved cooperation and shared responsibility. In fact, DevOps centers rapid, flexible development iterations, wherein chunks of code are produced and deployed independently and are supported by a high degree of automation.

All members of the studied teams were assigned to one of the two offices. We collected pseudo anonymized access card data to understand team behavior. The data contained information about employees' entry into and exit from office buildings, including timestamps and employee team affiliation. Using the team identifiers, we tracked the attendance rates and co-presence in different teams over time. Data over the period from January 2022 to October 2022 were obtained. We inspected the data for completeness and found missing data for some days across the sample period. To address this issue, we only considered the weeks with complete data for all workdays (Monday–Friday). Furthermore, as we study hybrid working, we removed the periods affected by the coronavirus disease 2019 (COVID-19) pandemic and by summer vacation. Accordingly, only the data from April, May, and October 2022 were retained. To preserve confidentiality, teams with fewer than five members were excluded. Data were extracted and analyzed using R open-source software.

We consider multiple variables related to office presence, which are defined below:

**Attendance rates:** For each day, we calculated the attendance rate as the number of employees that swipe their access card at any of the office locations divided by the total number of employees. We then computed weekly and monthly attendance as mean values over the days in a week and a month, respectively. Both the overall attendance rate for the 17 teams combined and the team attendance rate for each team were calculated. We also calculated individual attendance rates by dividing the days an individual member was at the office by the total number of working days in the sample period.

**Co-presence:** Similar to the time-overlap and time-separation configurations studied by Espinosa and Carmel (2003), we studied team co-presence patterns. For each dyad of team members, we calculated a co-presence index based on the proportion of workdays with overlapping office presence against all workdays in the sample period. Overlap occurs when both members are present at the same office during the workday. We calculated the team-level co-presence index for a team as the overlap of office presence per working day divided by the number of team member dyads. This index was 1 if each team member was co-located with all the other team members every workday.

**Coordination coefficient:** For each dyad of team members, we calculated a coordination coefficient by rescaling the co-presence index so that a value of 1 corresponds to the maximum possible co-presence, given the dyad's attendance rates, and a value of 0 corresponds to the expected co-presence index if the dyad's presence was independent. Similarly, for each team, we calculated a corresponding team coordination coefficient by rescaling the co-presence index so that a value of 1 corresponds to the maximum possible co-presence, given the team's attendance rate, and a value 0 corresponds to the expected co-presence index if the team members' presence was independent.

Further demographic analysis was performed for a subset of employees who had registered information about gender, age, and tenure in the company. We used this information as explanatory variables. We classified employees into four seniority groups: employees with less than 1 year of tenure, with 1–3 years of tenure, with 3–5 years of tenure, and with more than 5 years of tenure. Three years roughly corresponds to the median tenure in the dataset. Similarly, we classified employees into four age groups: younger than 30 years, 30– years, 40–49 years, and older than 50 years; the cutoff at 40 years roughly corresponds to the median age in the dataset.

# 4. Results

In this section, we report team attendance rate and co-presence and investigate the characteristics of employees with higher co-presence than others based on the factors affecting the coordination coefficient.

## 4.1 Attendance rates by teams

Office presence has been reported as a challenge in the post-pandemic period. Here, we report attendance at the team level—that is, the proportion of team members present at the office during the week. We calculated the average weekly attendance rate for each of the 17 teams (see Figure 1).

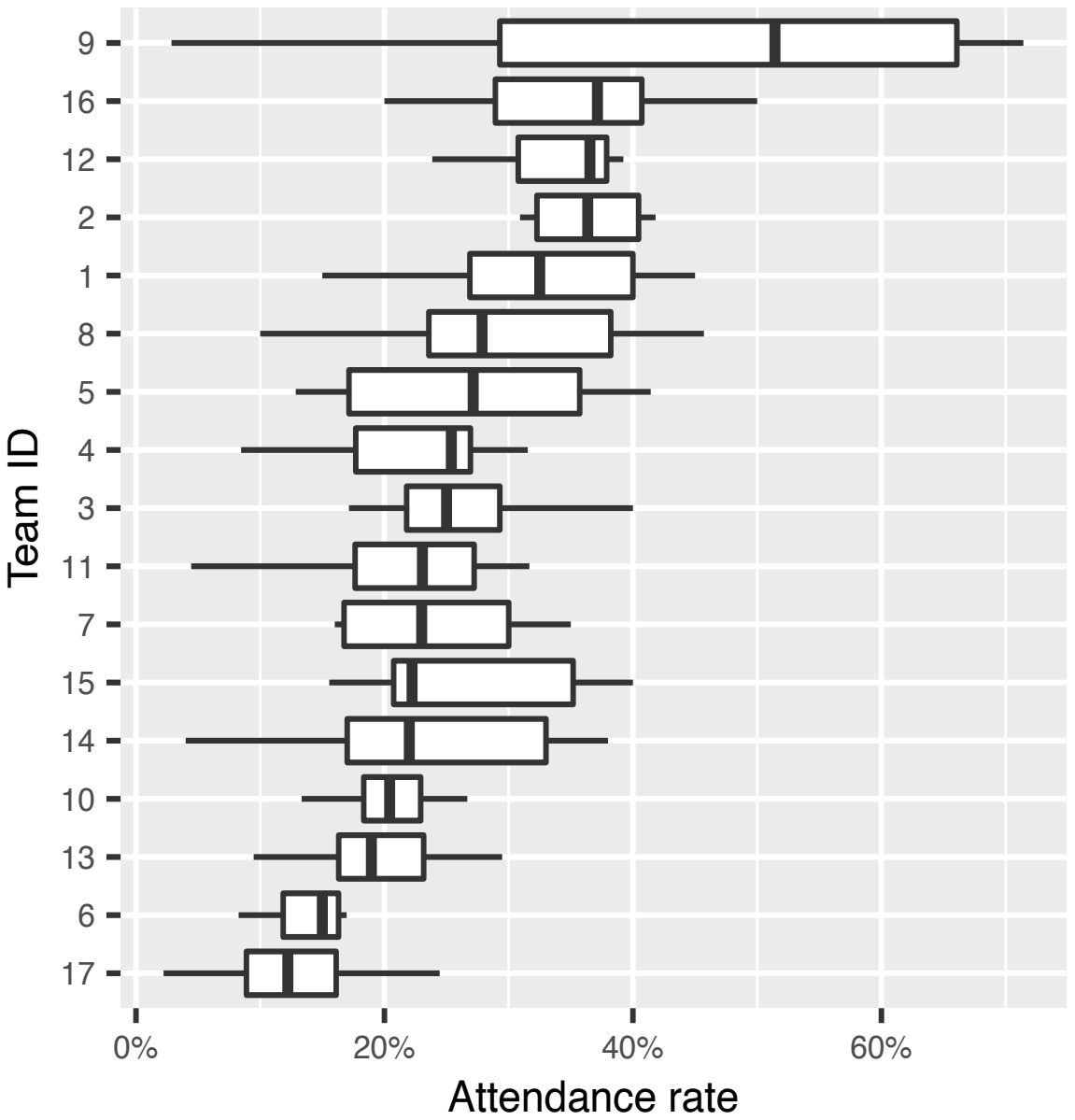

*Figure 1: Team-level attendance rates.*

The boxplot captures the distribution of the weekly attendance rates for each team. We observe a wide variation in the median attendance rates, ranging from slightly above 50% (half of the team members visiting the office during a week) to below 15% (few team members visiting the office during a week; the actual number depends on the team size). Evidently, no teams worked fully onsite or were fully remote. However, the lower whisker values suggest that four teams (team 9, 11, 14, and 17) have weeks where they work fully remote.

To understand what characterizes the members who are more frequently present at the office than others, we further analyzed the distribution of attendance rates over the sample period, stratified by gender, age, and tenure (Figure 2).

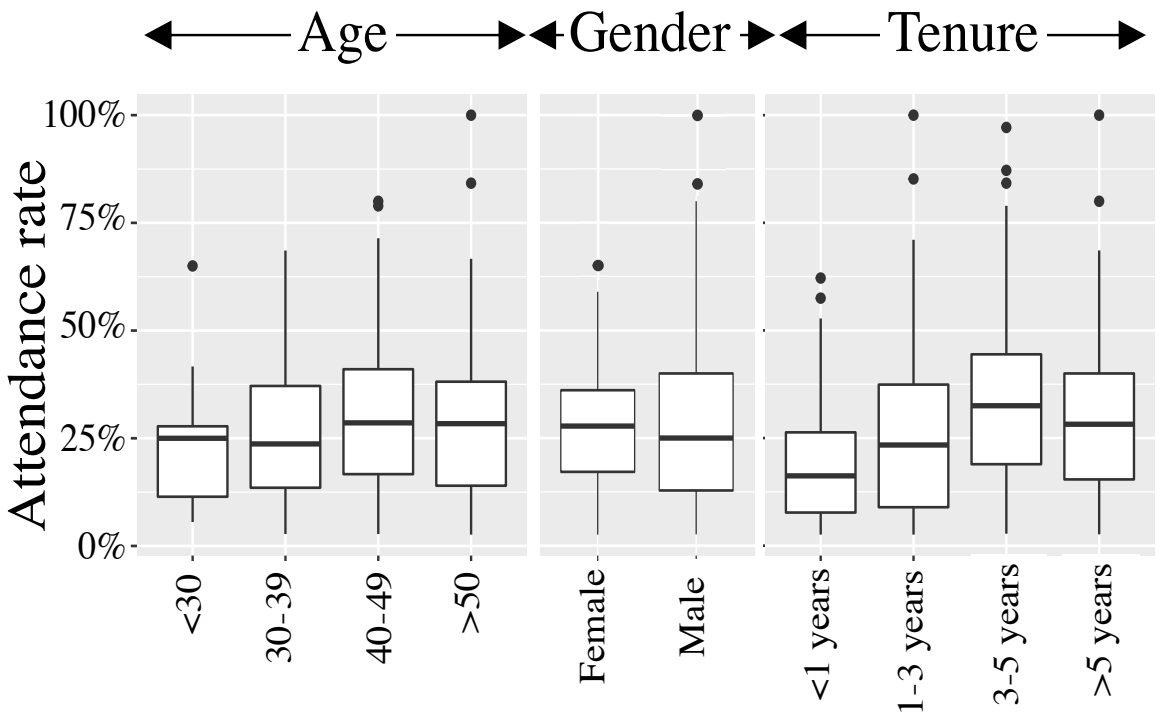

*Figure 2: Attendance rates for age, gender, and tenure.*

The average individual attendance rates for age and gender are around 25%. Older employees (older than 40) have slightly higher attendance rates than the younger ones, employees below 30 have less variation in attendance rates than those above 30, and females have slightly higher average attendance rates but less variation than males. Further, large differences in attendance rates exist between the employees who have been affiliated with the company for a long period (above 3 years of tenure) and relatively recent hires (below 3 years of tenure), which vary from 16% among new hires (<1 year) to 33% among those with 3–5 years of tenure.

## 4.2 Team co-presence

Team co-presence can determine the effectiveness of team communication and coordination as well as team psychological safety (Espinosa et al., 2007; Tkalich et al., 2022). Thus, focusing on co-presence is perhaps even more important than the attendance rate. The level of co-presence is a function of both how often team members work from the office and how coordinated they are in their attendance.

Team co-presence in our study comprises dyad co-presence. We define two team members as co-present on a given day if they are both present at the same office on that day. Over a specific period, we define the co-presence index for every dyad of team members in a team as the proportion of days in the period both members are at the office. This index ranges between 0 (they are both at the office on none of the days) and 1 (they are both at the same office every day). We now more closely examine the four teams that had different patterns of co-presence (one team from each of the quadrants in Figure 8).

Figure 3 illustrates one of the teams (team ID 2) as a network of employees, with the size of the edges representing the dyads co-presence index. The team has nine members and a mean weekly team attendance rate of 33%.

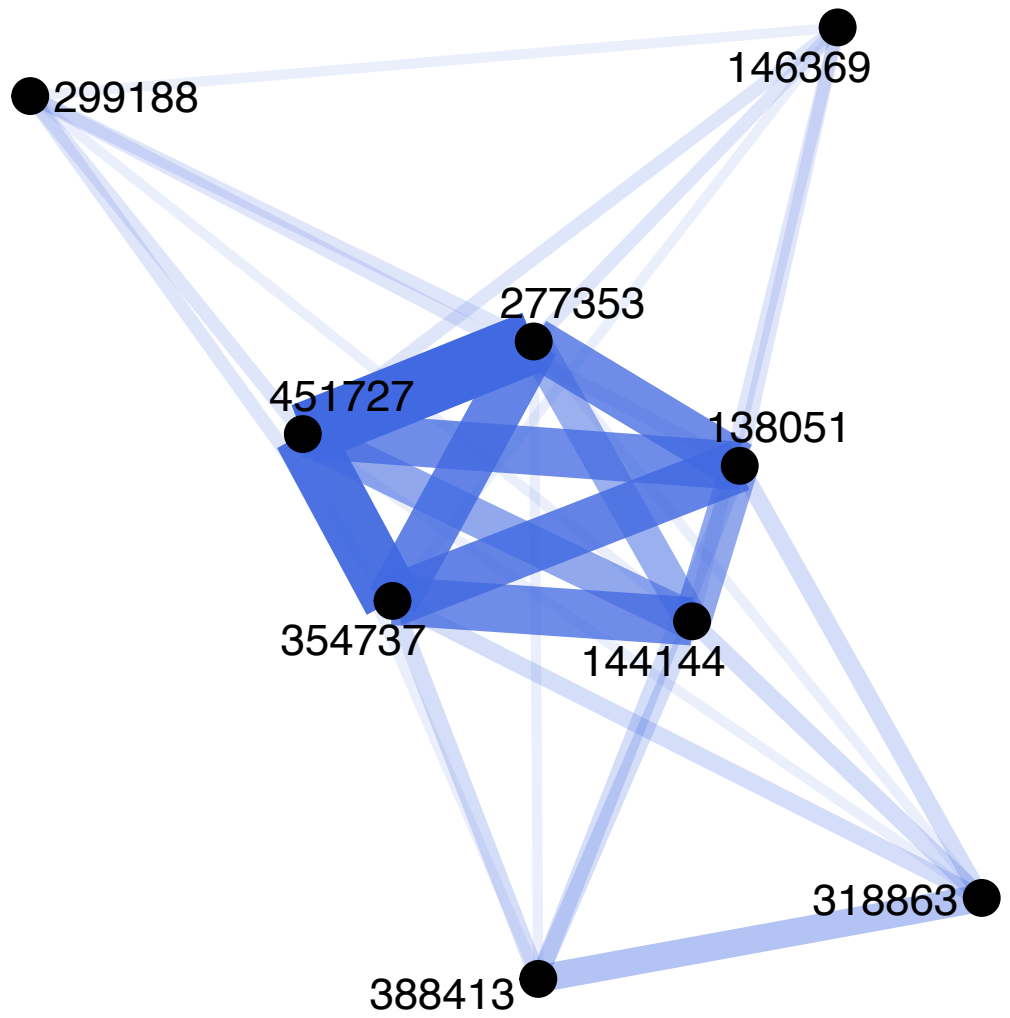

*Figure 3: Network of member co-presence in team 2.*

The team network (Figure 3) reveals some interesting patterns. For instance, a core group of five members seems to have significantly higher co-presence than the rest of the team. This team has a relatively high mean attendance rate (Figure 2); however, as apparent from Figure 3, several team members are rarely co-present in the office. Another interesting observation is that employees 388413 and 318863 are co-present more often with each other than with the other members, indicating the possibility of office attendance "shifts."

Figure 4 illustrates another example of a team (team ID 13) with 18 members and a relatively low weekly mean team attendance rate of 16% (Figure 2).

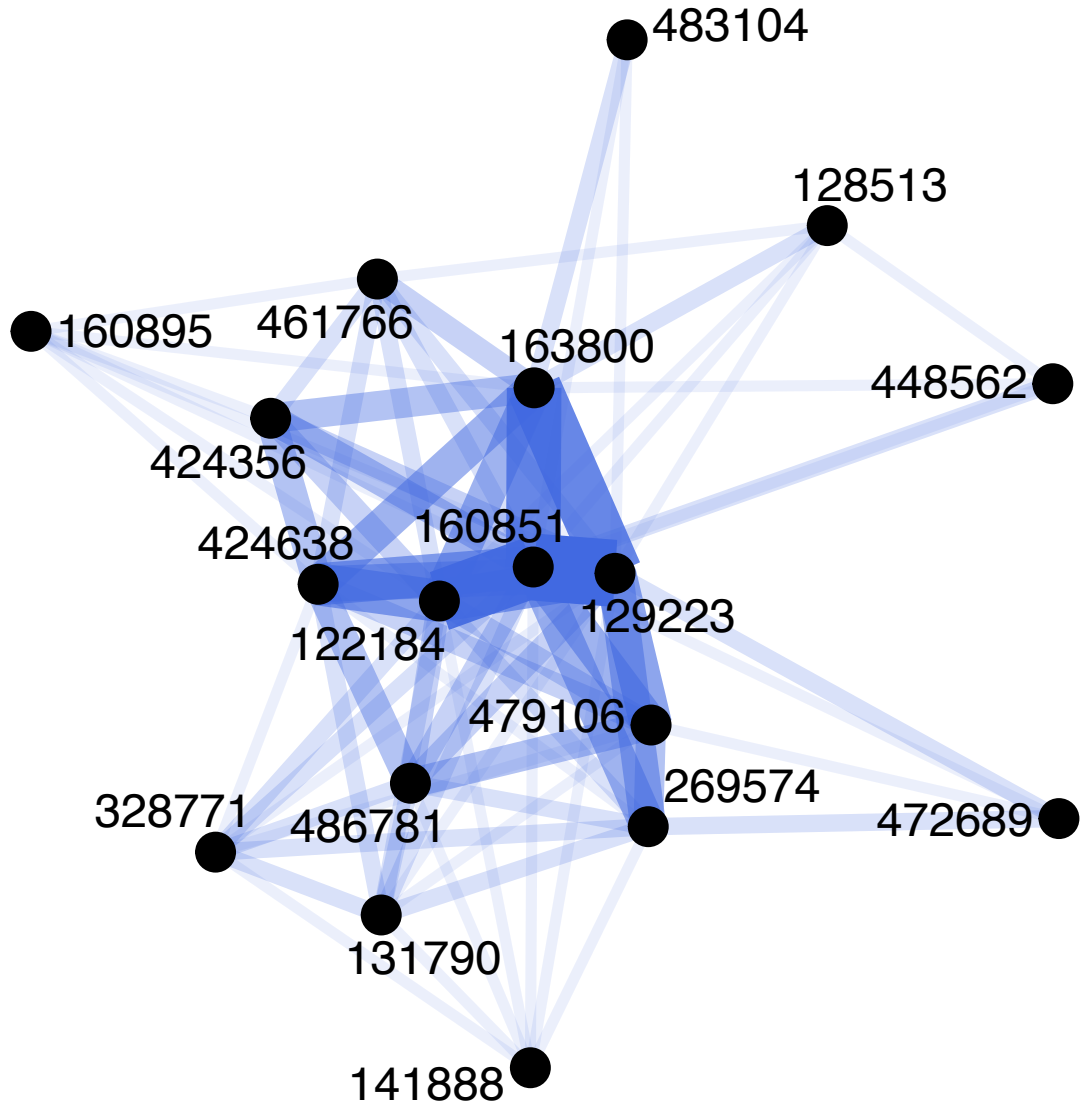

*Figure 4: Network of member co-presence of team 13.*

Several team members of team 13 have very low co-presence with other team members, indicating that many of them work mostly remotely. We do, however, note a group of team members with higher co-presence indices than others.

Figure 5 shows team 9, which has the highest attendance rate (Figure 2). Interestingly, in this large team, one member (398771) seems to be seldom present with the others, indicating that this member mostly works from elsewhere, as the team is frequently in the office.

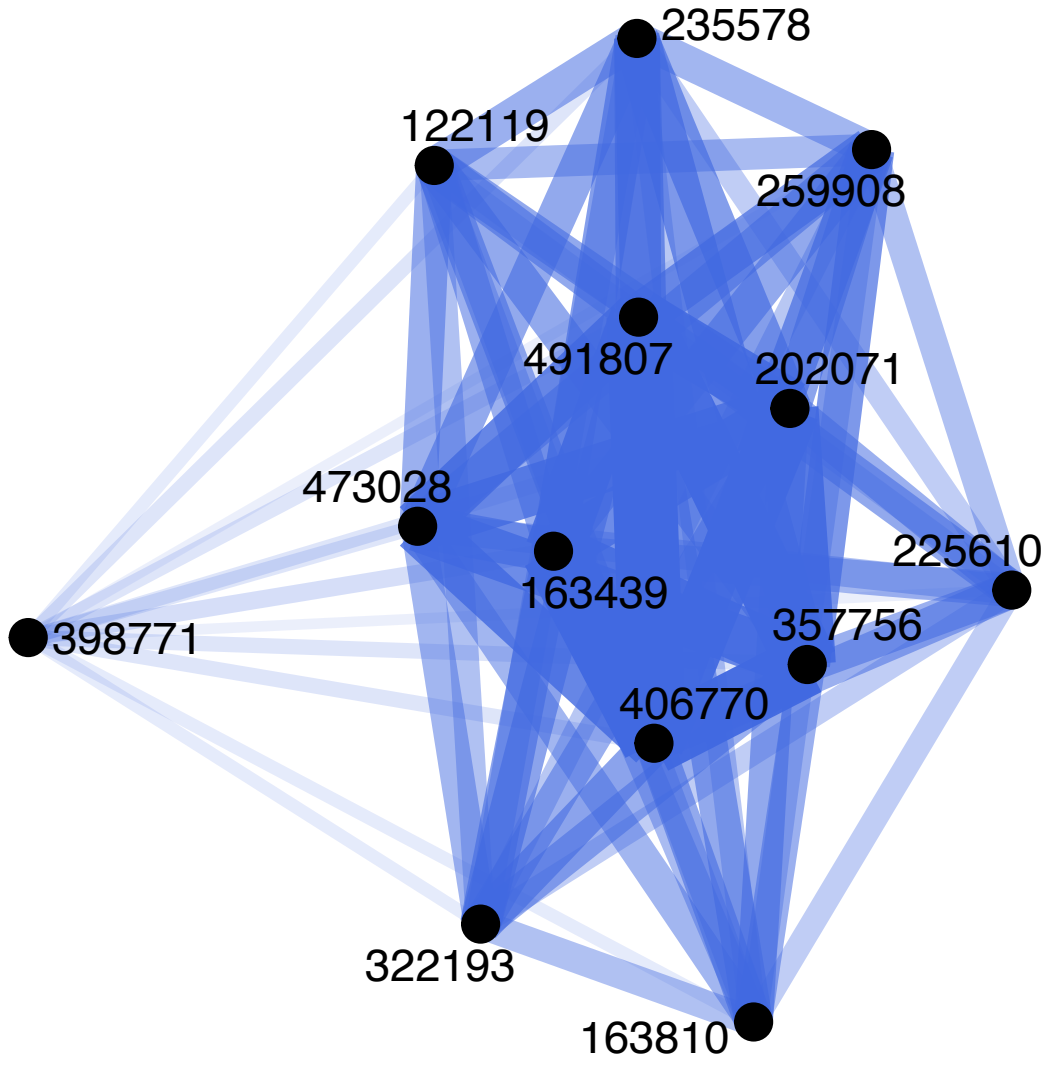

*Figure 5: Network of member co-presence in team 9.*

Figure 6 shows the network of member co-presence in team 14. This small team has a low attendance rate, and they work almost fully remotely in some weeks. Of the five team members, one is excluded from the graph, as this team member is working from another office. One member (108556) is seldom together with the others at the office.

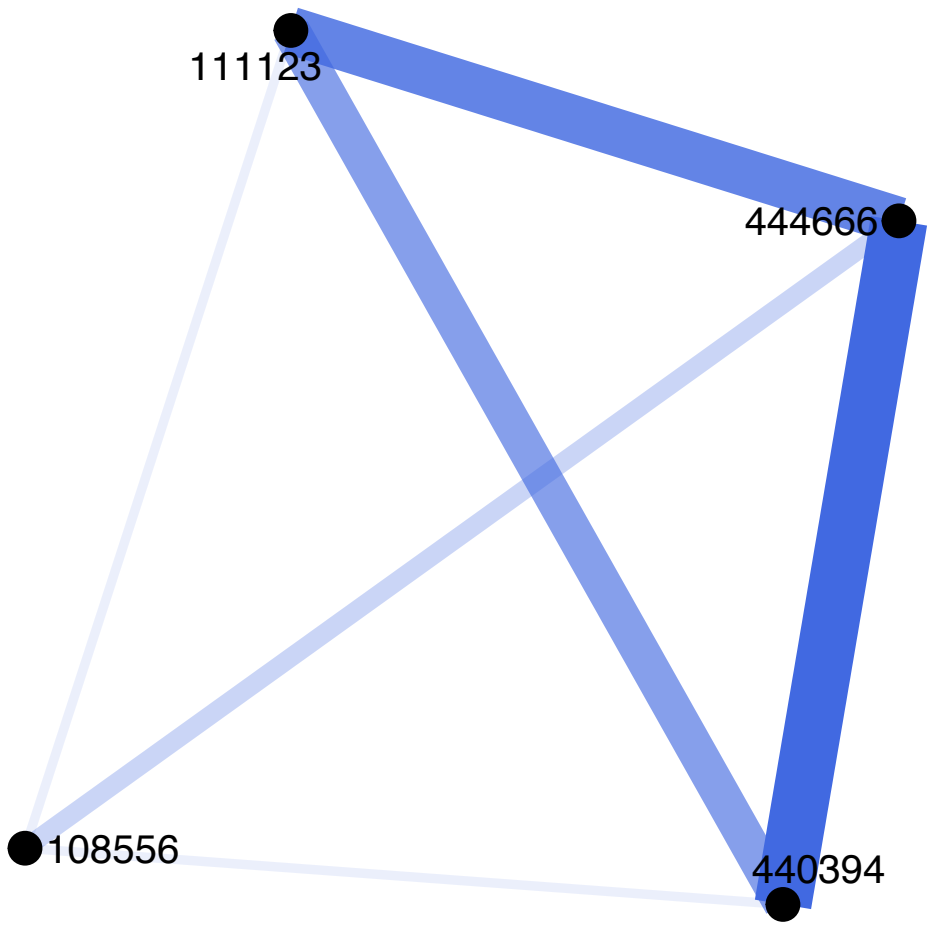

*Figure 6: Network of member co-presence in team 14.*

Similar to the attendance rates, we investigated the characteristics of team members who tend to have higher co-presence than others. Although we cannot claim with certainty that co-presence is coordinated in advance, our findings reveal interesting patterns.

For each dyad of team members, we calculated a coordination coefficient by rescaling the co-presence index so that a value of 1 corresponds to the maximum possible co-presence, given the team's attendance rate, and a value of 0 corresponds to the expected co-presence index if the team members' presence was independent.

To understand which factors affect team members' co-presence at the office, we applied an ordinary least squares (OLS) model, where the dependent variable is the coordination coefficient of a dyad of team members, and the explanatory variables are indicator variables capturing whether the dyad is of the same gender, whether they have similar tenures (e.g., both have tenures of less than 1 year), or whether they are in the same age group (e.g., both are below 30). Table 1 presents the results.

Table 1: Factors affecting the coordination coefficient

| **Characteristic** | **Coordination coefficient** |
|---|---|
| Same age | −0.023 (0.029) |
| Same gender | $0.050^{+}$ (0.029) |
| Same tenure | 0.073* (0.029) |
| Number of Obsservations. | 677 |

$^{+}$ $p < .1$, * $p < .05$, ** $p < .01$, and *** $p < .001$

We note that dyads of the same gender tend to work at the office on the same days more often than dyads of different genders. We also note that dyads of the same age group tend to be more co-present than dyads of different age groups, but the statistical significance of this relationship is weaker. Finally, no statistically significant effect of the tenure was observed.

## 4.2 Coordinated co-presence

In addition to mere co-presence, we investigated co-presence that does not occur by chance but is likely to be coordinated—that is, team members who have likely agreed to work at the office on the same days. To this end, we first calculated the team-level co-presence index for team $t$, denoted as $cp_t$ over the course of a period. This index is defined as the number of team interactions per day (denoted as $I_t$) divided by the number of dyads of team members, which for a team of size $N_t$ is given by $N(N-1)$.

$$cp_t = \frac{\bar{I}_t}{N_t(N_t - 1)}$$

If all team members are present at the office each day (so that all team members can possibly interact with each other each day), the co-presence index is 1. If no team members are present at the office at the same day, the index is 0. If we interpret the team as a weighted network, where the weight of the connection between each member is their co-presence index (as in Figures 3–6 above), the team-level index is equal to the density of this network. As illustrated in Figure 7, a team's co-presence index is highly correlated with its mean attendance rate. However, we also note that no direct relationship exists between team co-presence and mean attendance rate—if teams can coordinate their office presence, they can generate more co-presence for a given attendance rate (e.g., team 14).

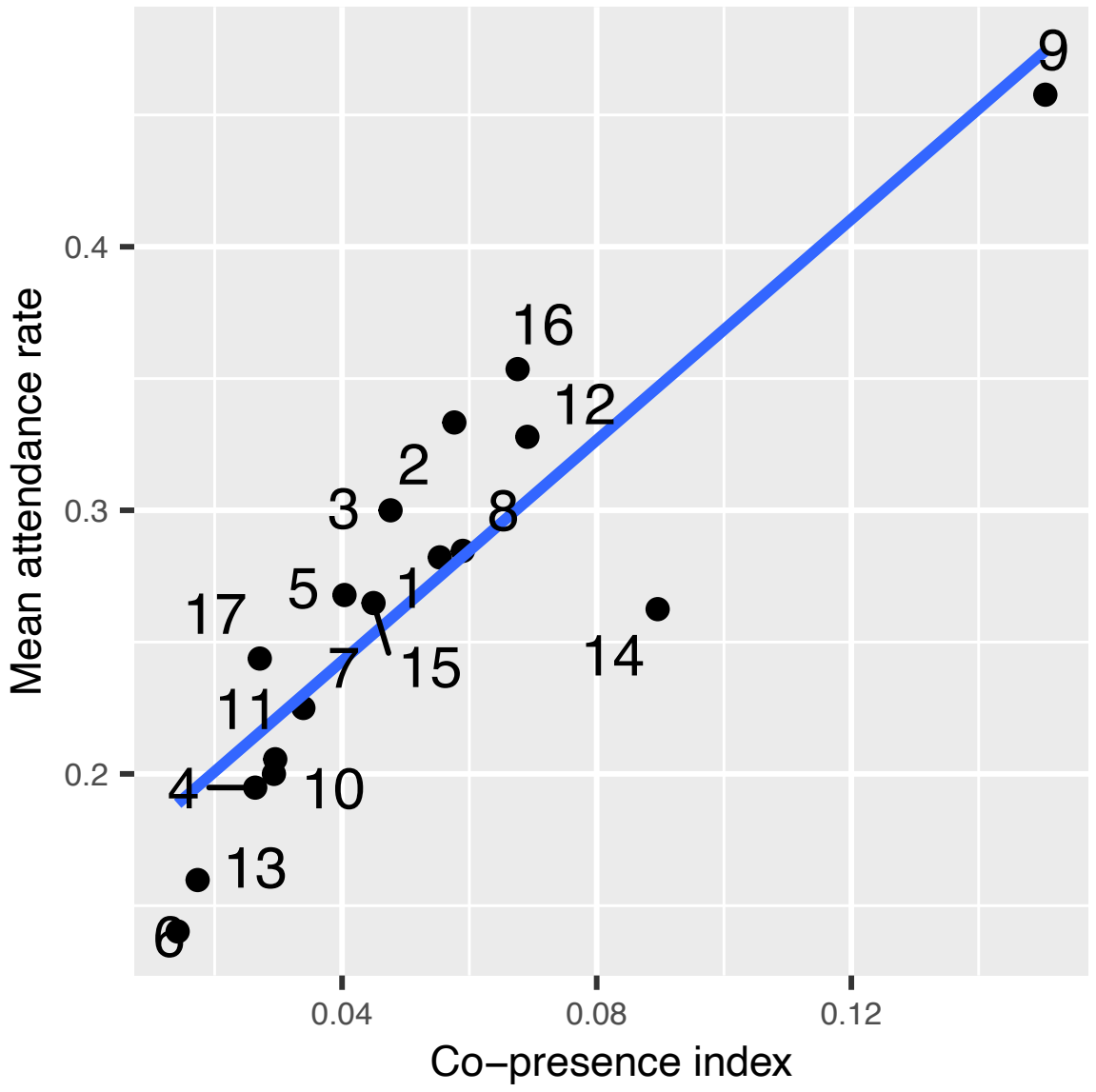

*Figure 4: Co-presence index and mean attendance rate. The blue line corresponds to a line fit by OLS.*

As office presence is a measure of how coordinated a team is, we define a coordination coefficient by rescaling the co-presence index so that a value of 1 corresponds to the maximum possible co-presence, given the team's attendance rate, and a value of 0 corresponds to the expected co-presence index if the team members' presence was independent.

In Figure 8, the coordination coefficient is presented on the x-axis and the mean attendance rate on the y-axis. Each point represents a team, and the team size reflects the team's co-presence index. We have also plotted a horizontal line at the median attendance rate and a vertical line at the median coordination coefficient. This allows us to characterize each team:

1) Teams located in the bottom left quadrant have low attendance rates and low coordination coefficients. One example is team 13 (Figure 4), where only half of the team is present with each other, forming a sub-team.
2) Teams in the bottom right quadrant are coordinated but have low attendance rates.
3) Teams in the top left quadrant have high attendance rates but are uncoordinated. One example is team 2 (Figure 3), which also reveals a pattern where 50% of the team is at the office together.
4) Teams in the top right quadrant have high attendance rates and are coordinated. One example is team 9 (Figure 5), where everyone except for one person is frequently at the office together. Team 14 is the most coordinated team and has an average office co-presence.

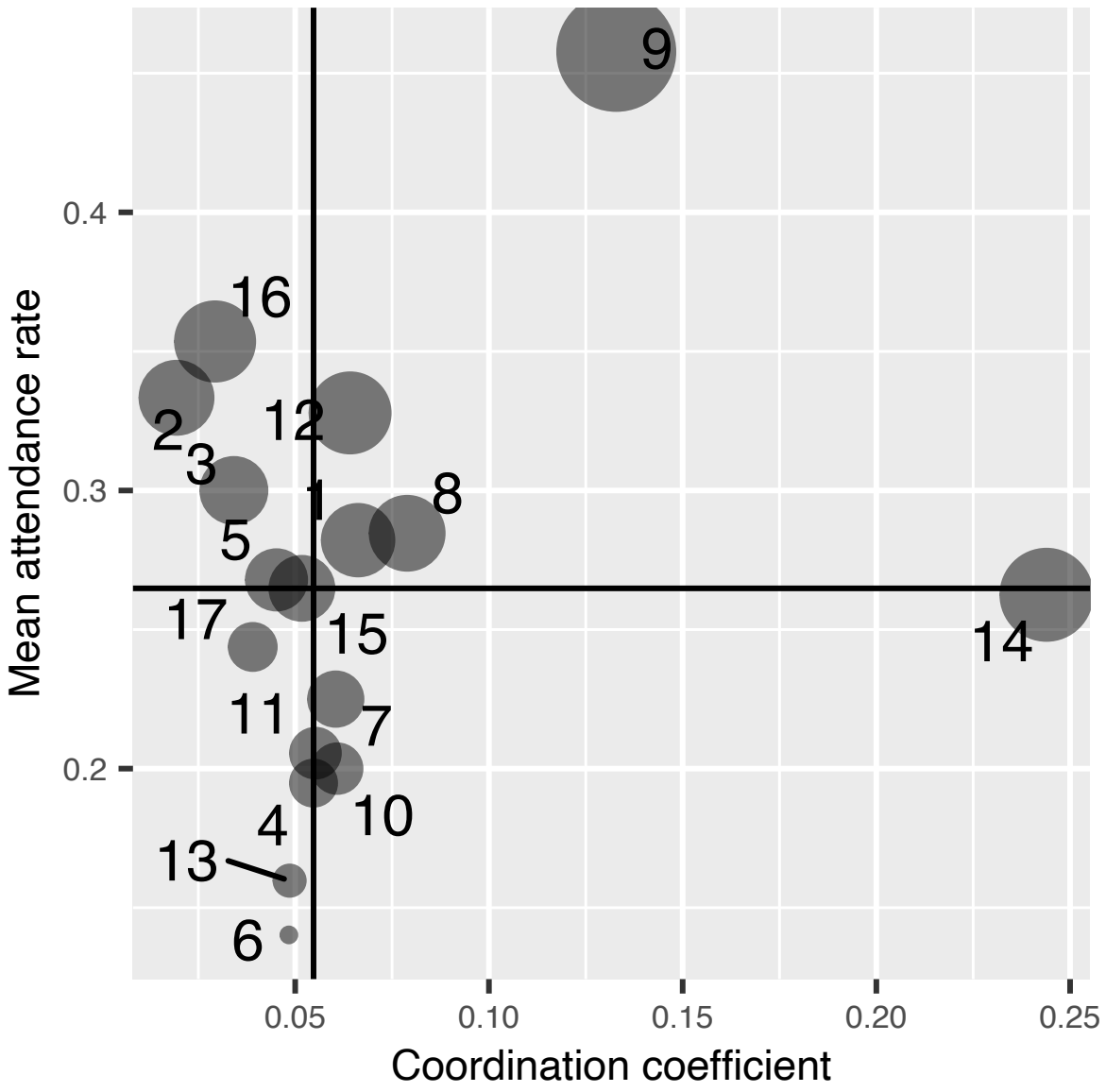

*Figure 5: The extent of coordination among teams.*

## 5. Discussion and conclusion

Recent studies have revealed that many companies are concerned that without employees' physical presence at the office, innovation, creativity, competence development, and knowledge sharing are likely to decline (Clear, 2021; de Souza Santos & Ralph, 2022; Yang et al., 2022). A key reason of this is the challenges in spontaneous interactions in a team when some team members are working from home. Further, psychological safety, company culture, the sense of belonging, attachment, and team cohesion are likely to suffer when teams do not coordinate their office presence (Tkalich et al., 2022).

By analyzing access card data, we investigated office presence among 17 teams in an agile telecommunications company over a period of 10 months. Below we describe our findings in light of our research question:

*How do agile team members organize their work rhythm in a hybrid work environment*?

**Summary of the findings:** Our results show that co-presence significantly varies among teams and team members. Further, age and tenure appear to influence office presence, and older employees prefer to work from the office more often than younger employees. Our findings are in line with the multiple case study performed by Smite et al. (2023), which revealed that older employees preferred working from the office and that gender differences do not affect office presence.

We found that some teams are coordinated, whereas others are not. Past research suggests that when teams coordinate their office presence, there is a positive effect on psychological safety (Tkalich et al., 2022), networking, and the onboarding process, as onboarding is easier when the new hire is in the office together with others. Thus, discussing office presence is key for all teams. Interestingly, we found that the most coordinated team had one of the lowest office presences (team 13). Further, the team with the fourth highest office presence (team 2) was the least coordinated team.

This finding extends existing research on the role of office presence, showing that high average office presence does not necessarily imply that team members are at the office on the same days. On the contrary, in some teams with higher office presence, team members seldom actually met while in the office.

When investigating what characterizes employees that tend to be co-present, we found that team members are likely to coordinate their office presence with members of the same gender and age group. This behavior raises serious concerns about the effectiveness of member onboarding in the team, especially because our results show that newly recruited people are more likely to be at home than others and less likely to be in the office with the seniors.

A key contribution of the above analysis is the finding that high average office presence in the team does not necessarily imply that team members meet often in person at the office. Our findings show that non-coordinated teams can have both high average office presence and low frequency of physical interaction among team members. Further, such teams seem to be divided in sub-groups, where about half of the members meet regularly. This is important because team leaders who assume that high average office presence indicates frequent personal contact may easily overestimate the actual level of personal contact.

**Implications:** Our findings have several important implications. First, research on onboarding has documented the importance of newcomers having access to seniors and mentors. Mentors can teach newcomers about the company, provide advice, and help with job instruction (Britto et al., 2017). Bauer (2010) points out that the opportunity to attend informal meetings with colleagues helps new employees more easily adapt to the new job. However, previous research has shown that spontaneous interactions are challenging in hybrid software development settings (Sporsem et al., 2022; Tkalich et al., 2022).

Therefore, teams that have members of different age groups, genders, and tenures and teams that form sub-groups need increased attention, especially when new team members are onboarded.

Many companies want to understand the best practices for institutionalizing hybrid work: how many days employees should be allowed to work from home and how many shall be spent in the office (Conboy et al., 2023). Our study adds a new, important question: How many days a team should spend together? Further, within the same organization, different teams perform different tasks, and people have different preferences regarding flexibility. Therefore, we conclude that there is no one way of hybrid working, as adjustments to the people, tasks, and characteristics of the team are required. If new people are onboarded, they need access to experts and must get to know everyone in the team. Such teams are found in the upper right quadrant of Figure 8. Also, teams that have a higher age group and tenure and are smaller in size are more coordinated. From this, we argue that agile teams in hybrid software development should be as small as possible (small is beautiful (Carmel & Bird, 1997)).

**Future work:** The effects of WFH have been found to make collaboration networks more heavily siloed across organizational units as well as more static with fewer new ties (Yang et al., 2022). Therefore, future work should investigate co-presence beyond the boundaries of the team—for example, on the same floor or in the same building—and whether co-presence has any influence on communication structures. Further, future work should investigate employee turnover and employee satisfaction surveys to better understand the effects of team co-presence.

**Limitations:** An apparent limitation of this study is that it is based on data from a single organization, which limits the generalizability of the findings. Further, we have only explored the data for when people are at work; this does not necessarily correlate with availability for interaction. People might be isolated in online meetings. People also might attend office for a half day and not meet their team members. Moreover, people could only come to the office to meet the access card swiping criteria. We also assume that the likelihood of co-presence beyond the expected is the result of team-member–coordinated office days, which might not be the case.

## Acknowledgements

This research is funded by the Research Council of Norway through the 10xTeams project (grant 309344) and the Swedish Knowledge Foundation through the KK-Hög project WorkFlex (grant 2022/0047).